# Low power, chip-based stimulated Brillouin scattering microwave photonic filter with ultrahigh selectivity

David Marpaung[1]*, Blair Morrison[1], Mattia Pagani[1], Ravi Pant[1#], Duk-Yong Choi[2], Barry Luther-Davies[2], Steve J. Madden[2], and Benjamin J. Eggleton[1]

[1]Centre for Ultrahigh bandwidth Devices for Optical Systems (CUDOS), the Institute of Photonics and Optical Sciences (IPOS), School of Physics, University of Sydney, NSW 2006, Australia

[2]Centre for Ultrahigh bandwidth Devices for Optical Systems (CUDOS), Laser Physics Centre, Australian National University, ACT 0200, Australia

*To whom correspondence should be addressed; email: d.marpaung@physics.usyd.edu.au

#Present address: Indian Institute of Science Education and Research, College of Engineering Trivandrum Campus, Trivandrum-695016 Kerala, India.

Highly selective and reconfigurable microwave filters are of great importance in radio-frequency signal processing. Microwave photonic (MWP) filters are of particular interest, as they offer flexible reconfiguration and an order of magnitude higher frequency tuning range than electronic filters. However, all MWP filters to date have been limited by trade-offs between key parameters such as tuning range, resolution, and suppression. This problem is exacerbated in the case of integrated MWP filters, blocking the path to compact, high performance filters. Here we show the first chip-based MWP band-stop filter with ultra-high suppression, high resolution in the MHz range, and 0-30 GHz frequency tuning. This record performance was achieved using an ultra-low Brillouin gain from a compact photonic chip and a novel approach of optical resonance-assisted RF signal cancellation. The results point to new ways of creating energy-efficient and reconfigurable integrated MWP signal processors for wireless communications and defence applications.

The explosive growth in mobile communications demands radio-frequency (RF) technologies with exceptional spectral efficiency such as cognitive radios, which can adapt their frequencies to exploit the available spectrum in real-time [1,2]. Such frequency-agile systems will benefit hugely from RF filters that can be tuned over many gigahertz whilst keeping high MHz-scale resolution and high selectivity to prevent severe interference due to spectrum-sharing. While this is difficult to achieve with all-electronic filters [3-7], integrated microwave photonic (IMWP) filters [8] can readily achieve multi-gigahertz tuning range without significant degradation in their frequency response. However, these filters typically exhibit limited resolution (GHz instead of MHz linewidths) and are plagued by trade-offs between key parameters, such as between the frequency tuning range and the resolution for multi-tap filters [9-13]; or between the peak rejection and the resolution for resonator-based filters [14-18].

Stimulated Brillouin scattering (SBS) [19-22] offers a route to MHz-resolution IMWP filters. Although SBS has been widely studied in optical fibers, recently there has been a growing interest in harnessing SBS in nanophotonic waveguides [22-27]. The ability to control the coherent interaction of photons and acoustic phonons in chip-sized devices (as opposed to in optical fibers many kilometres long) promises not only fascinating new physical insights, but also opens the path to realising key technologies on-chip including slow light [28,29]; narrow linewidth lasers [30]; optical frequency combs [31,32]; RF signal processing [33-35] and filtering [36-40]. In particular, SBS filters can exhibit linewidths of the order of 10-100 MHz. Such a high resolution is unmatched by most on-chip devices because it requires extremely low material losses and impractically-large devices [41].

Although IMWP filters exploiting SBS on chip have been demonstrated, they need relatively high SBS gain and high pump powers to achieve moderate filter suppression. In [40], Morrison *et al* demonstrated a bandstop filter with 20 dB suppression in a 6.5 cm $As_2S_3$ waveguide using 350 mW of pump power. From a broader perspective, this need for high power is unattractive for on-chip signal processing with particularly stringent requirements for energy efficiency. More importantly, this prevents implementation of integrated RF signal processing in low gain SBS devices, such as the ones recently reported in CMOS-compatible platforms [26,27]. By harnessing SBS in silicon, these devices are highly attractive due to the potential to integrate RF signal processing in a single compact monolithic chip [42]. Nevertheless, they currently exhibit very low gain, of the order of 1-4 dB, which is far from sufficient for any MWP signal processing if one relies on conventional techniques.

Here, we demonstrate experimentally a highly selective SBS IMWP bandstop filter in a cm-scale chalcogenide glass waveguide that operates with a low pump power (8-12 mW) and a low SBS gain (1-4 dB), while maintaining high, reconfigurable resolution (32-88 MHz) and high stop-band rejection of >55 dB. We further show that the filter can be tuned over a wide frequency range of 0-30 GHz, leading to a unique performance combination difficult to match with any existing filter technology. We achieved this performance through on-chip SBS filtering of a phase and amplitude-tailored RF-modulated optical spectrum [43-45], resulting in precise RF signal cancellation at a resolution comparable to state-of-the-art RF filters. We further show that for a given SBS gain, this approach allows the flexibility to re-distribute the pump optical power into modulated optical power, thereby reducing the filter insertion loss. The results presented here point to new possibilities for creating high performance SBS-based reconfigurable MWP filters that will play a key role in modern RF systems for next generation radar [46] and high data rate wireless communication [47], with a potential for monolithic integration in silicon chips (Figure 1a).

# Results

**Enhancing filter suppression by spectral tailoring**

The chip-based SBS filter topology is illustrated in Figure 1b. An electro-optic modulator is used create RF-sidebands from an optical carrier. The modulated signal is injected into a nonlinear chalcogenide optical waveguide [48,49] as the probe of an SBS process, while an SBS pump is injected from the opposite end. Our filter relies on precise phase-and amplitude tailoring of the RF-modulated optical spectrum at the input of the SBS filter (Figure 1c). The desired spectrum contains a dual-sideband modulation (DSB) where the optical sidebands are out-of-phase but have unequal amplitudes. The SBS gain spectrum is used to amplify the weaker sideband to achieve the same amplitude as the stronger sideband only at the peak of the gain resonance. The sidebands along with the optical carrier are then sent into a high speed photodetector for direct detection that results in RF mixing between these components. At the RF frequency where maximum sideband amplification occurs, the mixing products between sidebands and the optical carrier have equal amplitudes but opposite phase and thereby cancel leading to very high signal suppression [43,44]. Outside the SBS resonance, the signals do not completely cancel due to the amplitude difference between the sidebands [50]. This filtering technique contrasts with the conventional SSB approach (Figure 1d) where the SBS loss/absorption resonance is used to attenuate the sideband in a single sideband (SSB) optical spectrum. Whilst simple, this approach suffers from a lack of suppression and low resolution due to inherent limitations in the SBS loss resonance [44].

It is important to stress that the tailored spectrum required for the filter operation is very different from spectra generated and used in conventional MWP signal processing so far such as intensity modulation (IM), phase modulation (PM), or SSB modulation. Ideally, the spectrum should contain an optical carrier and sidebands that can be tailored independently in amplitude and phase. A good approximation of this ideal spectrum can be generated using a dual-parallel Mach-Zehnder modulator (DPMZM). In the small signal approximation this electro-optic modulator can create an optical carrier with two sidebands whose relative amplitude and phase difference can be sufficiently tunable [42,50].

**On-chip filter experiments**

Figure 2a depicts the experimental setup to demonstrate the filter operation. An SBS pump at 1550 nm was generated, amplified and injected into a 6.5 cm-long $As_2S_3$ optical waveguide via a lens-tipped optical fiber. A frequency-detuned RF-modulated probe generated using a DPMZM was launched from the opposite end. The typical insertion loss of the photonic chip was measured to be 9.5dB. The waveguide exhibited a large Brillouin gain coefficient, $g_B = 0.74\times10^{-9}$ m/W due to its high acoustic confinement and small effective area [23] (see Methods for detail of the device fabrication). At the pump side, an optical circulator was used to collect the transmitted filtered optical signal. Photodetection and RF mixing occurred in a high speed photodetector. The filter response was measured using a vector network analyzer (VNA). Details of the experimental setup can be found in the Methods section.

Using the experiment setup we demonstrated and compared the filtering performance of the conventional SSB approach and the novel cancellation-based filter. The conventional filter generated a SSB RF modulated optical spectrum as the input to the optical filter. Typically, the measured extinction ratio of the suppressed optical sideband with respect to the unsuppressed sideband was 20 dB (Figure 2b). The cancellation filter requires a near phase-modulation signal as input (see Figure 1c). The measured optical spectrum of the generated sidebands with opposite phase and 1 dB amplitude difference is shown Figure 2c.

Figure 2d shows the measured RF magnitude responses of both the conventional and the cancellation approaches. In the conventional approach, a bandstop filter with moderately high suppression (20 dB) can only be achieved by pumping the optical waveguide with a high pump power (350 mW), as shown as the blue trace of Figure 2d. When very low pump power, in the order of 8 mW, was used instead, the filter suppression was very shallow, of the order of 0.8 dB (green dashed trace). On the other hand, with the same low pump power, and a sub-1 dB SBS gain, the cancelation filter can achieve 55 dB of suppression and a higher resolution compared to the conventional filter (thick red trace). This result illustrates the superiority of the novel filter, which exhibited impressive peak suppression and a high resolution whilst using only low SBS gain and very low pump power. This massive 43-fold reduction in required pump power (8 mW versus 350 mW) highlights the energy efficiency enhancement of the novel cancellation filter.

**Frequency tuning and bandwidth reconfigurability**

We demonstrate the frequency agility of the MWP filter by tuning the central frequency of the RF bandstop response and its 3-dB bandwidth. Central frequency tuning with a resolution of 12.5 MHz was achieved by adjusting the frequency difference between the pump and the probe waves by temperature tuning the probe DFB laser. For all measurements, the pump wavelength and power at the facet of the photonic chip were kept at 1550.43 nm and 24 dBm, respectively. The chip insertion loss was 8.5 dB. The result is depicted in Figure 3a, where central frequency tuning from 1 to 30 GHz was achieved. Critically, the filter suppression was maintained above 51 dB for the entire tuning range.

Reconfiguration of the filter bandwidth was achieved by switching between SBS gain and loss responses, and the variation of pump power for each case. The RF bandwidth of the bandstop filter is equivalent to that of the SBS process, given by [50]

$$\Delta \upsilon_{RF} = \Gamma_B \sqrt{\frac{G}{\ln(e^G + 1) - \ln 2} - 1} \quad (1)$$

where $\Gamma_B$ is the Brillouin linewidth, and the $G$ parameter is proportional to the SBS pump power (see Supplementary material). Here, $G > 0$ corresponds to SBS gain, while $G < 0$ corresponds to SBS loss. Figure 3b shows the calculated filter bandwidth as a function of $G$, for $\Gamma_B = 58$ MHz, together with the experimental data. For the SBS gain, we tuned the SBS pump to achieve gain variation from 0.8 dB to 11.6 dB and obtained increase in filter resolution from 56 MHz down to 32 MHz. As for the SBS loss, we tuned the peak absorption from -0.8 dB to -8.1 dB to expand the filter bandwidth from 61 MHz to 88 MHz. In total, we achieved 56 MHz tuning range in the experiment. For each measurement we kept the stopband suppression beyond 50 dB by changing the balance between the sidebands to maintain cancellation at the center of the SBS resonance. The normalized frequency response of the filter at the extreme points of the bandwidth tuning range is shown in Figure 3c.

**Demonstration of RF filtering**

Here we demonstrate what is to our knowledge the first high resolution RF filtering experiment using a chip-based MWP filter. We consider a scenario where two RF signals, one of interest and the other unwanted interferer, were supplied to the input of an optical modulator (see Figure 4). The signal of interest was band-limited with a width of 1 MHz, whilst the unwanted interfering signal was a single frequency tone. These were separated in frequency by 20 MHz, and the power of the unwanted signal was 23 dB higher than the signal of interest (see Methods for detail of RF signal generation and measurement). We compare the filtering performance of this input spectrum between a conventional SSB filter and the novel cancellation filter. The SSB filter was generated using SBS loss with 28 dBm pump power at the chip facet creating 17 dB of peak suppression. The cancellation filter was created using 4 dB of SBS gain from 21 dBm of pump power at the chip facet. The chip insertion loss in these measurements was 8.8 dB. In both measurements the filter response was aligned to have maximum suppression at the unwanted tone frequency of 12.002 GHz.

The measured output RF spectra with and without the SBS pump in the conventional SSB filter are depicted in Figure 4a. As expected the unwanted tone power was reduced by 17 dB, but the signal attenuation was as high as 9 dB, which indicated that the conventional filter resolution was below 20 MHz. This clearly demonstrates the limitation of the conventional approach which cannot simultaneously achieve high resolution and high suppression filtering. In contrast, this can be achieved using the cancellation filter, as shown in Figure 4b. The measured interferer suppression in this case was 47 dB, limited by the noise floor of the measurements. The signal underwent a low attenuation of 2 dB, indicating that the half-width at half-maximum of the filter was below 20 MHz.

**Insertion loss reduction experiment**

For various reasons, such as photosensitivity [48], heating [39], or intensity-dependent losses [27], many on chip devices are not suited to handle high optical power. These devices are thus expected to work with a stringent optical power budget. Here, we investigate and compare the overall performance of the conventional SSB filter and the cancellation filter under a total optical power budget of the order of 400 mW at the facets of the chip. Such a power was chosen to optimise long term operation and stability of the chip based filter. We quantify the filter performance in terms of peak suppression, resolution, and RF-to-RF filter insertion loss.

In the case of the conventional SSB filter we distributed the optical power as follows: 25 dBm (316 mW) as the SBS pump and 20 dBm (100 mW) as the (probe) modulated optical carrier power. Higher pump power was chosen to maximise the suppression of this filter. The insertion loss of the photonic chip in these experiments was 9.3 dB. We generated an RF filter with 13 dB suppression and a resolution of 100 MHz. The measured RF insertion loss was -37.7 dB, obtained using 3 mA of detected photocurrent. The measured response centered at frequency of 24.5 GHz is shown in Figure 5 (blue trace). We then reversed the optical power distribution between the pump and the probe in the case of the cancellation filter. With 100 mW of pump power we generated 4 dB of SBS gain. The filter response exhibited 55 dB of peak suppression and a resolution of 40 MHz. The detected photocurrent increased to 10 mA due to higher optical carrier power, and an insertion loss of -31.3 dB was measured which is a 6.3 dB improvement compared to the case of the conventional SSB filter.

The results of these experiments demonstrate that in the case of limited power budget, the cancellation filter allows re-distribution of optical power from pump to probe waves, thereby reducing the insertion loss of the filter, with improved peak suppression and resolution compared to the conventional filtering approach. Combined with a higher power handling photonic chip and a photodetector that can handle higher photocurrent, the cancellation filter can potentially achieve very low insertion loss in addition to the high suppression, high resolution, and wide frequency range – a combination very difficult to achieve using any existing RF or MWP filtering approach.

## Discussion

We compared the overall performance of the SBS-on-chip cancellation filter to other state-of-the-art integrated filter technologies. This is summarized in Table 1. First, the SBS-based filter uniquely combines salient features of two different MWP filter classes; high rejection levels typical of a multi-tap filter, and wide frequency tuning typical of resonance-based filters. In terms of linewidth, our filter is unmatched by any other integrated MWP approach, achieving nearly two orders of magnitude higher resolution. In fact, such a high resolution is more akin to low-loss electronic filters [3-7]. However, for the same resolution and rejection level, the SBS filter achieves 36 times wider tuning range compared to the best performing electronic RF filter reported very recently [7]. Overall, our filter showed the highest quality factor (Q = 375) and the fractional tuning range (2900%) compared to any integrated filter technology to date.

The lowest insertion loss measured in our filter was -30 dB, which is lower than typical losses in integrated MWP filters [11,12] but is high compared to electronics solutions. This loss can be recuperated using two approaches: reducing chip-coupling loss and reducing the loss in the underlying photonic link. Lower chip coupling loss, of the order of 1.1 dB per facet, can be achieved in optical waveguides with inverse tapers [51]. Photonic link loss can be minimized by using an RF modulator with lower half-wave voltage ($V_\pi$), higher power laser source, and a higher power handling detector [52,53]. Realistically, the insertion loss figure can be improved by 28 dB using the waveguide with inverse tapers, a modulator with half the $V_\pi$, and a high photodetector current of 40 mA. This will lead to -2 dB insertion loss which is comparable to electronic RF filters, and is crucial to realize a filter with low noise figure and high spurious-free dynamic range (SFDR) [54,55]. Such an improvement in the insertion loss will lead to a filter technology with all-optimized properties that is essential in modern RF systems and applications.

We note that the approach introduced in this paper can be generalised to schemes that use SBS for band pass filters, such as the recent demonstrations reported in [37, 39, 56, 57], offering better extinction and more energy efficient operation in photonic integration platforms.

## Methods

**Device fabrication**

The waveguide was fabricated by thermally evaporating a 0.85 micron thick film of $As_2S_3$ (refractive index n = 2.43 at 1550 nm wavelength) on an oxidized silicon wafer. From this, straight rib-type waveguides of width 4 µm were patterned using standard contact photolithography. Surrounding trenches were then dry-etched 380 nm deep using inductively coupled plasma reactive ion etching. After removing the photoresist, the chip was overclad with a UV cured polysiloxane resin (n = 1.53 at 1,550nm). Its end facets were prepared by hand-cleaving the silicon substrate using a diamond scriber. Anti-reflection coating was applied at both facets. The rib waveguide has a length of 6.5 cm and cross-sectional area of 4 µm x 0.85 µm.

**Experimental setup**

For the SBS pump we used a DFB laser (Teraxion Pure Spectrum λ = 1550 nm) with 100 mW of optical power, connected to an electronically-controlled variable optical attenuator (VOA, Kotura), and an EDFA (Amonics) with maximum output power of 1 W. For the SBS probe we used a DFB laser (Teraxion) to generate the optical carrier which was modulated using a DPMZM (EOSPACE IQ-0D6V-35). The modulator optical input and output were pigtailed using polarization-maintaining fibers. The half-wave voltage, insertion loss, and 3 dB bandwidth of the modulator were 5 V, 5 dB, and 35 GHz, respectively. The RF input ports of the DPMZM were driven by an RF signal through a quadrature hybrid coupler with frequency range of 1.7-36 GHz (Krytar). The three bias voltages of the DPMZM were adjusted using a programmable multi-channel voltage supply with 1 mV voltage accuracy (Hameg HM7044G). Optical coupling into the waveguides was through a pair of lensed fibres with spot diameter of 2.5 μm. The polarization of the light going into the photonic chip was controlled using fiber polarization controllers (PCs). The filtered optical spectrum was collected from the pump side through an optical circulator. The signal was detected using a high-speed photodetector (PD, u2t XPDV2120) with 0.6 A/W responsivity and 50 GHz RF bandwidth. For filter response measurements, a frequency-swept RF signal with 0 dBm power was supplied from and measured on a 10 MHz-43.5GHz vector network analyzer (VNA, Agilent PNA 5224A). Measurements of the modulated optical spectra were done using an optical spectrum analyzer (ANDO AQ6317B) with a resolution of 0.01 nm.

**RF filtering measurement**

The composite of RF signals used in the demonstration of RF filtering were generated as follows: an RF tone with a frequency of 11.982 GHz and power of − 9 dBm was generated using an RF synthesizer (HP 8673H). The tone was frequency modulated with unity modulation index by a triangular waveform with 30 kHz frequency and 1 V peak-to-peak amplitude using an arbitrary waveform generator (Tabor Electronics WW5061). This created a band-limited signal with 1 MHz width and peak power of -21 dBm, used to emulate the signal of interest. The unwanted tone at 12.002 GHz and power of 0 dBm was generated using the signal source of an Agilent Fieldfox N9918A microwave handheld combination analyzer. The two RF signals were combined and supplied to an electro-optic modulator using a 1.7-36 GHz microwave combiner (Krytar). The RF spectrum analysis was done using the N9918A microwave combination analyzer with a span of 50 MHz and 100 Hz resolution bandwidth.

**Definition of filter Q-factor and fractional tuning range**

The Q-factor is defined as the ratio of the filter's 3-dB bandwidth to the central frequency of the resonance. The fractional tuning range (in percentage) is defined as the ratio of the frequency tuning range to the lowest center frequency of the stopband.

## Acknowledgement

Funding from the Australian Research council (ARC) through its Laureate Project FL120100029 and Future Fellowship FT110100853 are gratefully acknowledged. This research was also supported by the ARC Center of Excellence for Ultrahigh bandwidth Devices for Optical Systems (project number CE110001018). D. Marpaung and B. J. Eggleton acknowledge support from U.S. Department of the Air Force through AFOSR/AOARD (grant # FA2386-14-1-4030).

## Author contributions

D.M., B.M., and R.P. developed the basic idea for the ultra-high bandstop filter. D.M., B.M., and M.P. designed the experiments. B.M. and M.P. carried out the experiments with assistance of D.M. and R.P. M.P. carried out the numerical simulations. D.Y.C, S.J.M., and B.L.D. fabricated the device. D.M. wrote the manuscript. D.M. and B.J.E. supervised the project. All authors discussed and commented on the manuscript.

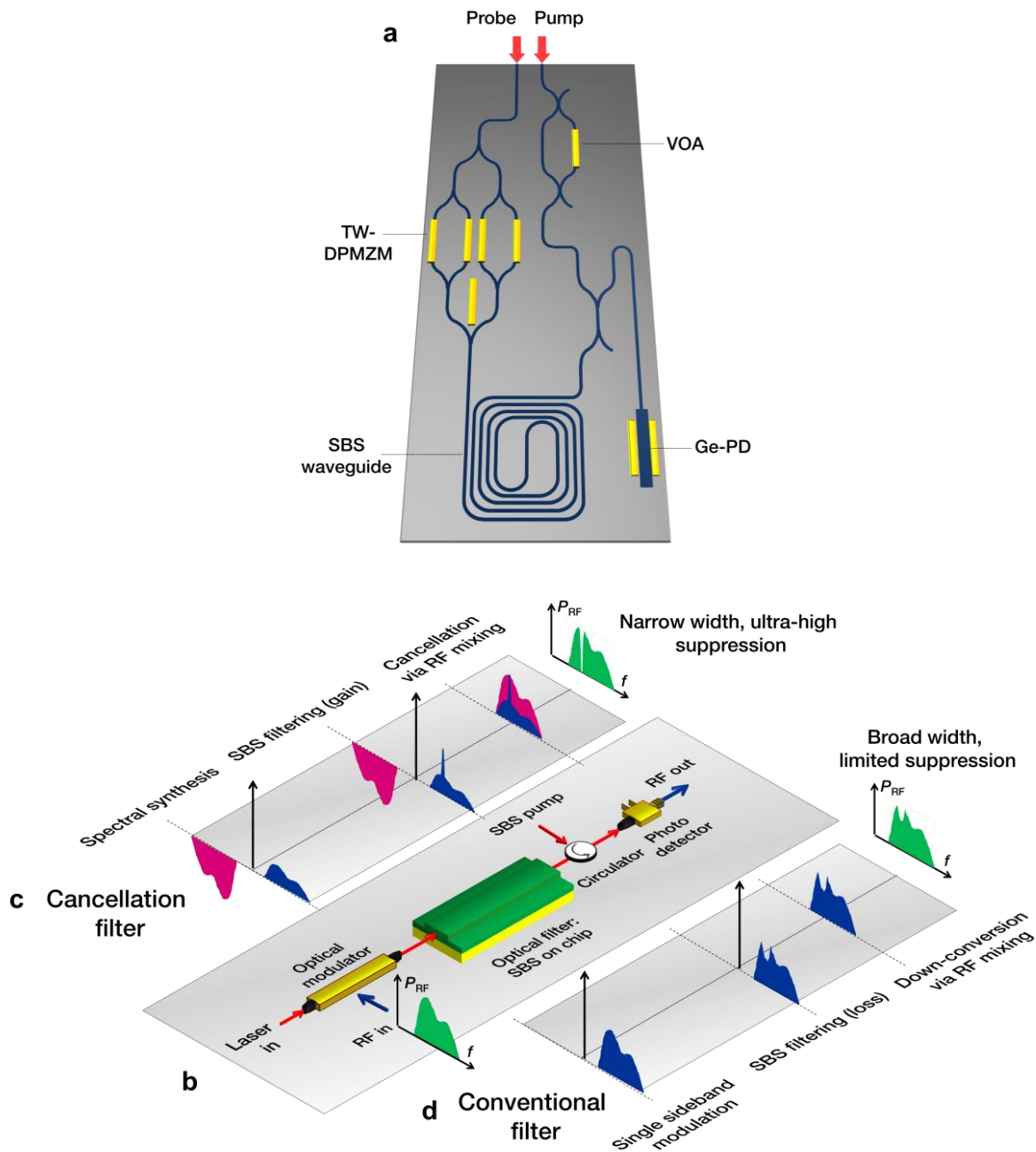

**Figure 1 | SBS-based integrated microwave photonic filter.** (**a**) Artist's impression of a future monolithic-integrated high suppression and reconfigurable SBS MWP filter in a silicon chip. VOA: variable optical attenuator, TW-DPMZM: travelling wave dual-parallel Mach-Zehnder modulator, Ge-PD: germanium high speed photodetector. (**b**) The schematic shows the topology of the microwave photonic filter reported here. An optical modulator was used for RF modulation spectral synthesis while stimulated Brillouin scattering in a chalcogenide waveguide was used as a reconfigurable optical filter. (**c**) In the novel cancellation filter nearphase-modulation signal (opposite-phase, unequal-amplitude sidebands) were generated and processed using SBS gain spectrum, leading to a highly selective filter. (**d**) In the conventional filter, single sideband spectrum was generated and processed using SBS loss/absorption spectrum, resulting in a filter with low selectivity.

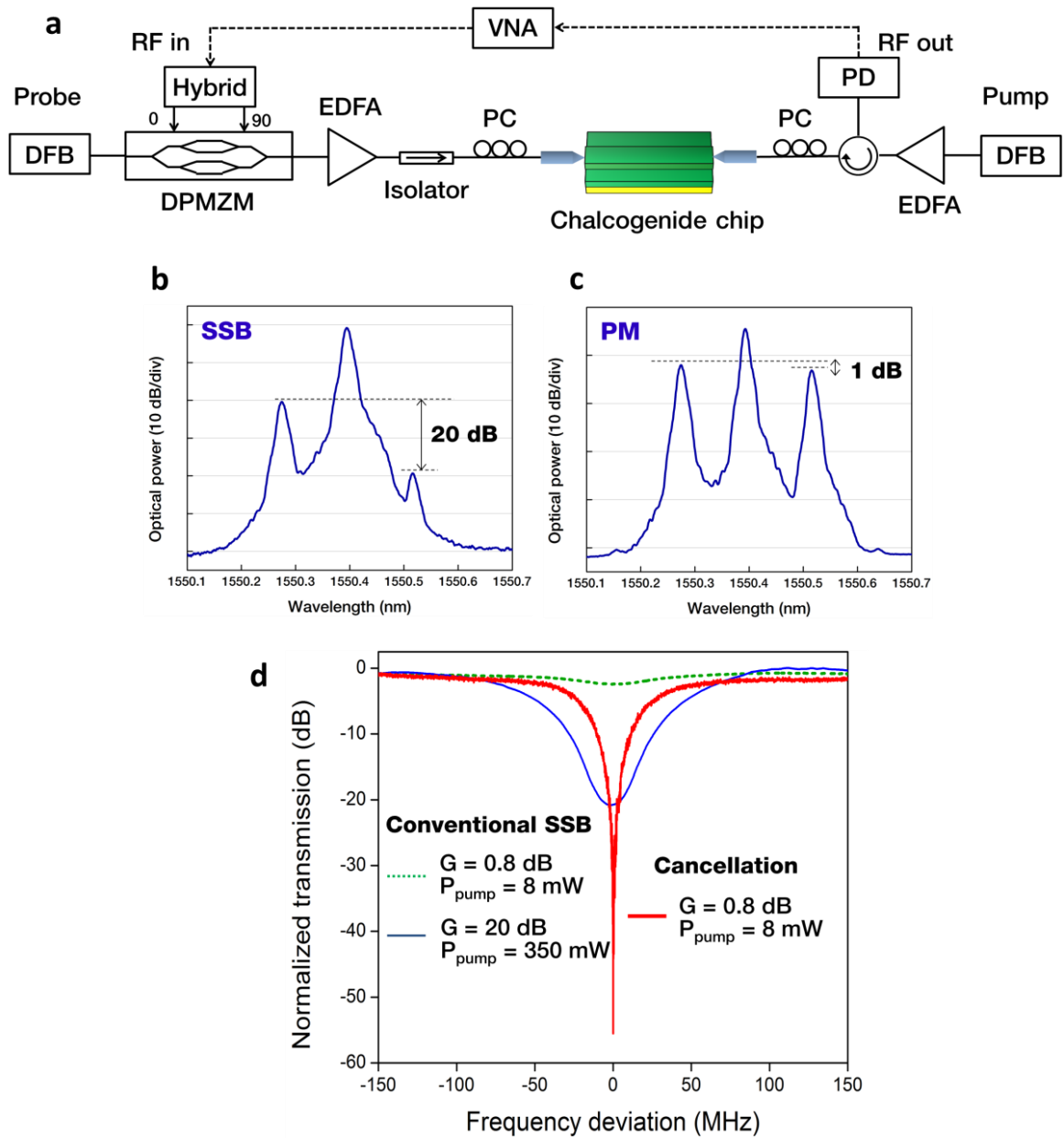

**Figure 2 | Microwave photonic filter experiments.** (**a**) Pump-probe experimental setup for filter performance evaluation, including Distributed Feedback Laser (DFB), 90° RF Hybrid Coupler (Hybrid), Dual-Parallel Mach-Zehnder Modulator (DPMZM), Erbium-Doped Fiber Amplifier (EDFA), Polarization Controller (PC), Photodetector (PD), and Vector Network Analyzer (VNA). (**b**) Optical spectrum measurement of input RF modulated signal for the conventional single sideband (SSB) filter. (**c**) Optical spectrum input for the cancellation filter, yielding near phase-modulation (PM) with unequal amplitude sidebands. (**d**) Corresponding VNA traces depicting filter responses for the conventional SSB and cancellation filters. For the same low pump power (8 mW), SSB filter yield 0.8 dB suppression, while cancellation filter yield 55 dB of suppression.

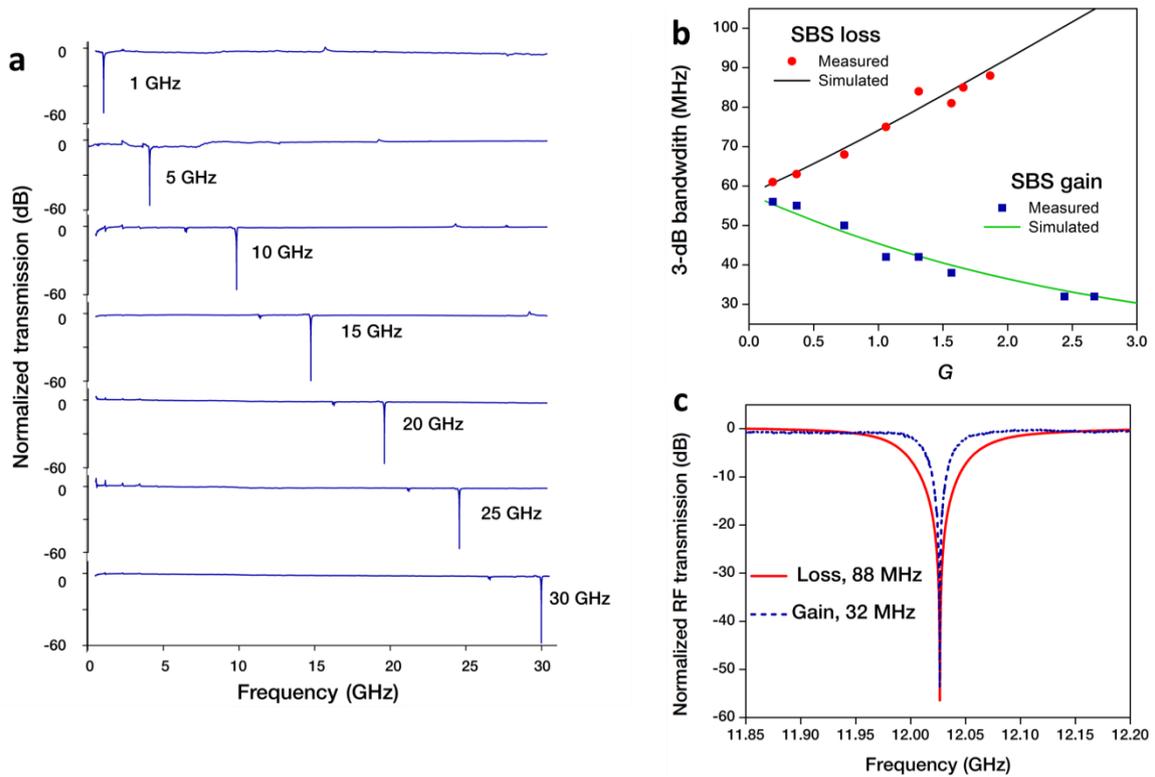

**Figure 3 | Frequency agility of the filter.** (**a**) Stopband center frequency tuning. Filter suppression was kept above 51 dB in all measurements. (**b**) Bandwidth tuning from 32 MHz to 88 MHz was achieved by means of tuning the pump power to vary SBS gain and loss. (**c**) Filter response at the extremes of the bandwidth tuning range.

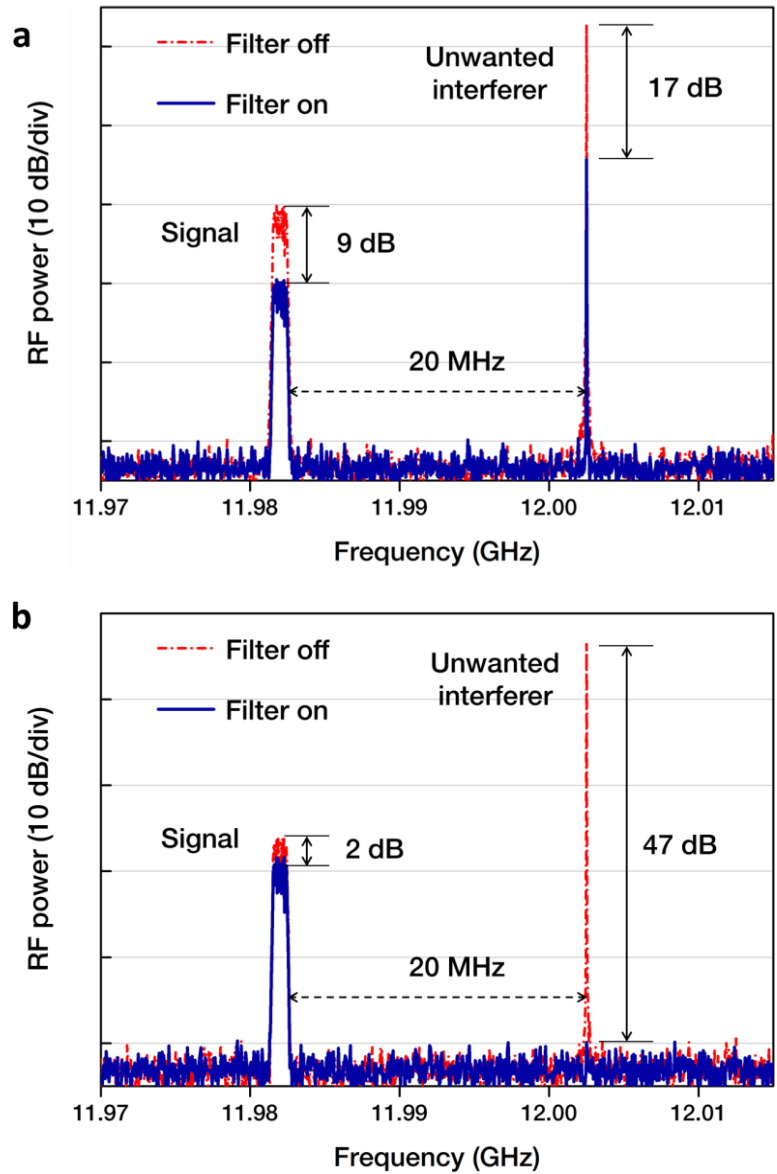

**Figure 4 | High resolution RF filtering experiment.** Two RF signals with 20 MHz frequency separation were used at the filter input. (**a**) Filtering with conventional single-sideband scheme with 17 dB SBS loss as optical filter. Peak attenuation at the unwanted interferer tone was 17 dB, and signal attenuation was 9 dB. (**b**) Filtering with the cancellation filter using 4 dB of SBS gain. Complete reduction of unwanted interferer was observed with low attenuation of the desired signal (2 dB).

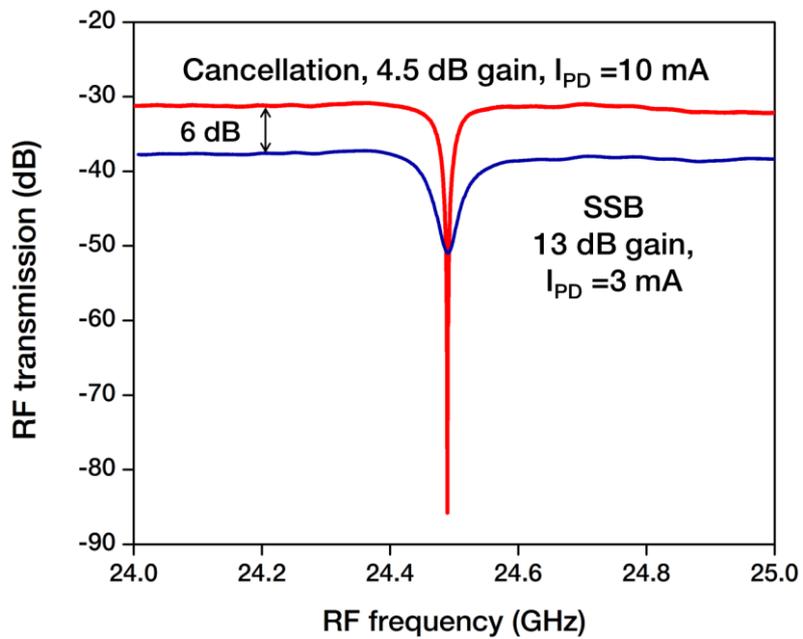

**Figure 5 | Filter insertion loss reduction.** Experiments for an optical power budget of 500 mW at the facets of the optical chip. Blue trace: Conventional single sideband approach with 25 dBm of input pump power and 20 dBm of probe power. Red trace: Cancellation approach with 20 dBm of pump power and 25 dBm of probe power. $I_{PD}$ : detected photocurrent.

**Table 1 | Performance comparison of state-of-the-art microwave band-stop filter technologies**.

| Technology | Class | Size | Line-width (MHz) | Freq. tuning range (GHz) | Rejection level (dB) | Insertion loss (dB) | Highest Q | Tuning range (%) |
|---|---|---|---|---|---|---|---|---|
| SBS on chip (this work) | MWP cancellation | cm | 33-88 | 1-30 | > 55 | -30 | 375 | 2900 |
| Photonic crystal [12] | MWP multi-tap | mm | 1000 | 10-30 | >50 | -45 | N/A | 200 |
| Frequency comb [11] | MWP multi-tap | km | 170 | 0-12 | >60 | -40 | N/A | N/A |
| $Si_3N_4$ ring [43] | MWP cancellation | mm | 247-850 | 2-8 | >60 | -30 | 28 | 300 |
| Silicon ring [16] | MWP single resonance | mm | 910 | 2-15 | >30 | N/A | 16 | 650 |
| Silicon ring [18] | MWP single resonance | mm | 6000 | 2.5-17.5 | >40 | N/A | 1.84 | 600 |
| MEMS tunable-absorptive notch [7] | Electronics | cm | 35/306 | 4-6/6.3-11.4 | >35 | -2 | 128 | 80 |
| MEMS tunable-resonator [3] | Electronics | cm | 77 | 8.92-11.34 | >24 | -0.8 | 90 | 27 |
| RF-CMOS resonators [6] | Electronics | μm | 710 | 2.9-4.3 | >50 | -1.95 | 6 | 48 |
| MEMS tunable-absorptive notch [5] | Electronics | cm | 9.7 | 3.4-3.8 | >30 | -0.3 | 345 | 11 |